# PRNU Based Source Camera Identification for Webcam Videos

Fernando Martín-Rodríguez.

fmartin@tsc.uvigo.es.

atlanTTic research center for Telecommunication Technologies, University of Vigo,
Campus Lagoas Marcosende S/N, 36310 Vigo, Spain.

*Abstract*- **This communication is about an application of image forensics where we use camera sensor fingerprints to identify source camera (SCI: Source Camera Identification) in webcam videos. Sensor or camera fingerprints are based on computing the intrinsic noise that is always present in this kind of sensors due to manufacturing imperfections. This is an unavoidable characteristic that links each sensor with its noise pattern. PRNU (Photo Response Non-Uniformity) has become the default technique to compute a camera fingerprint. There are many applications nowadays dealing with PRNU patterns for camera identification using still images. In this work we focus on video, more specifically on webcam video, because of the great importance of webcam video nowadays. Three possible methods for SCI are implemented and assessed in this work.**

## I. INTRODUCTION

Camera fingerprints are based on the unavoidable imperfections present in the image sensors due to their manufacture process. These imperfections derive from the different sensitivity to light of each individual pixel (hence the name, PRNU: Photo-Response Non Uniformity). Mathematically, this phenomenon is modeled as a multiplicative noise with zero mean that acts according to this equation [1], [2], [3]:

$$Im_{out} = (I_{ones} + Noise_{cam}) . Im_{in} + Noise_{add} \qquad (1)$$

Where $Im_{in}$ is the "true" image in front of the (incoming light intensity), $I_{ones}$ is a matrix with all values equal to 1 and $Noise_{cam}$ is the "sensor noise" (positive for pixels that are more sensitive than expected, negative in the opposite case). Symbol "**.**" means element-wise product and $Noise_{add}$ is additive noise from other source.

All SCI methods are based on somehow estimating $Noise_{cam}$ in order to compare results from different images.

This kind of methods is directly applicable to still images, in videos we can apply them defining some process that works with individual video frames.

The remainder of this text is organized as follows: first, we briefly describe the PRNU footprint and its calculation; then we describe the designed camera identification methods; finally, we describe the implementation of the tests and their results.

## II. PRNU: CAMERA FINGERPRINT

As already mentioned, PRNU is an acronym for Photo-Response Non-Uniformity and refers to the sensor noise defined in equation (1) [4]. The PRNU pattern has become the "de facto" standard for camera identification. PRNU estimation is always based on some type of image denoising filter to estimate $Im_{in}$ in equation (1). By having a set of images (which we know were captured by the same camera), a PRNU pattern can be estimated and then compared to any image from an unknown camera. The comparison should produce a high positive value if patterns are similar, which means that the source camera is likely to be the same. Today, despite recent advances in this field, it is very difficult to reach a legally irrefutable result.

The denoising process to compute PRNU can be from very different types, for example:

- A simple 3x3 median filter [5].
- The well-known Wiener filter [6].
- Wiener filter modifications, for example in [4] (the implementation that we use as basis), they use a WVT based filter (WVT: Wavelet Transform [7]). This means using Wiener equation ($H = \frac{Image}{Image+Noise}$) with WVT's. A constant WVT (white) noise is assumed.

Finally, it is important to use a good comparison method to detect similarities between the different patterns extracted. According to [4], we will base ourselves on PCE (Peak to Correlation Energy). This parameter is calculated by correlating two images (complete correlation for all possible displacements) and dividing the energy (squared value) of the maximum by the average energy calculated after "removing" the peak surroundings. PCE has shown good performance in the comparison of PRNU patterns [8].

## III. IDENTIFICATION METHODS

In all cases, one or more training videos per camera are selected. A sample rate is also initially defined that will determine the number of frames to be used. For a rate of 1/N, we will select one of each N "frames". With those images, the PRNU pattern for that sensor is obtained. The method is the one described in [4]: Wiener-type noise reduction is applied on wavelet transform and it is considered that the residual W (see equation 2):

$$W = Im - denoise(Im) \qquad (2)$$

Is equal to the product $Noise_{cam}Im_{in}$ (neglecting the additive noise, $Noise_{add}$). We also assume that $Im_{in}$=denoise(Im) and we compute the pattern (fingerprint) as the weighted average:

$$F = \frac{\sum W^n Im_{in}^n}{\sum (Im_{in}^n)^2} \qquad (3)$$

To identify the camera in a video of unknown origin, we have tried three different methods:
- Voting: identification is performed on each frame by maximum PCE as if they were individual images (again one image from each N). Each identification counts as "one vote". Simple majority wins.
- Pattern correlation: one of each N frames is extracted. With all of them, a new PRNU pattern is calculated that is compared with the previously stored ones. We have tried two different comparisons: a simple correlation and PCE. Results were very similar, slightly better for the first option. Working with patterns that have already been averaged, it seems that PCE does not add value.
- PCE vectors: for each frame, a vector is calculated with all the PCE values (one for each possible camera). These vectors are averaged, and the maximum value defines the final result. Two variants have been tested: normalizing each vector (dividing by its maximum) and without such normalization. The second option has worked better (note that the PCE parameter is already a normalized value).

## IV. TEST DESIGN

The tests have been carried out with a database made up videos captured by available cameras. These are 5 cameras, specifically: Conceptronic Amdis (low cost), Webcam integrated in HP laptop, Webcam integrated in HP All-in-One computer and Logitech C920 (two different cameras are available for this model). All videos have been obtained by recording directly in MPEG-4 (AVC: Advanced Video Coding) format.

We have performed three different tests:
- Test 1: the training is carried out with videos from the collection (without paying attention to the content).
- Test 2: training is carried out with specifically prepared videos. With simple content, constant color backgrounds: black, gray, white, red, green and blue (this is to avoid that a complex image complicates fingerprint extraction).
- Test 3: same as 2 but HD videos (1920x1080) are mixed with medium resolution ones (1280x720). Training has always been done in medium resolution. When finding a HD video, the frames are rescaled to the medium resolution size. We use the "nearest neighbor" option (without interpolation) to keep the original values.

For all cases, the confusion matrices defined as the number of times that the test samples of class i are identified as belonging to class j are calculated.

For each calculated matrix it is easy to calculate the success rate: $p = \frac{\sum_i CM_{ii}}{\sum_i \sum_i CM_{ij}} 100$, and also error rate: $q = 100 - p$.

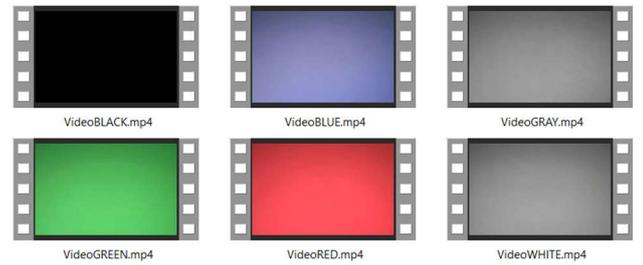

Fig. 1. Simple videos for training.

## V. TESTS RESULTS

All the tests have been carried out in the MATLAB environment [9]. The results for the three methods are compiled in the following tables where the percentage of error obtained for each test is shown. Three methods (one table per method) have been tested with different sampling rates. In each table, each line corresponds to a value of the sampling rate. It is ordered by increasing rate, in each line more frames are being used than in the previous ones.

TABLE I
Voting method

| Tasa | Test 1 (%) | Test 2 (%) | Test 3 (%) |
|---|---|---|---|
| 1/30 | 32.14 | 27.27 | 31.91 |
| 1/25 | 32.14 | 12.12 | 17.02 |
| 1/20 | 28.57 | 18.18 | 21.28 |
| 1/15 | 28.57 | 18.18 | 23.40 |
| 1/10 | 17.86 | 9.09 | 14.89 |

TABLE II
Simple correlation of PRNU patterns

| Tasa | Test 1 (%) | Test 2 (%) | Test 3 (%) |
|---|---|---|---|
| 1/30 | 28.57 | 21.21 | 27.66 |
| 1/25 | 39.29 | 24.24 | 27.66 |
| 1/20 | 25.00 | 12.12 | 17.02 |
| 1/15 | 28.57 | 15.15 | 21.28 |
| 1/10 | 28.57 | 9.09 | 12.77 |

TABLE III
PCE vectors

| Tasa | Test 1 (%) | Test 2 (%) | Test 3 (%) |
|---|---|---|---|
| 1/30 | 35.71 | 21.21 | 23.40 |
| 1/25 | 28.57 | 21.21 | 27.60 |
| 1/20 | 28.57 | 15.15 | 21.28 |
| 1/15 | 32.14 | 21.21 | 27.66 |
| 1/10 | 21.43 | 9.09 | 17.02 |

The conclusions drawn from these tables would be:
- Tests two and three always obtain better results, showing that training through specific videos works better.
- Test 3 always obtains worse results than test 2, showing that "extrapolating" the training to other resolutions is not entirely reliable. Here we really do not know very well if the relationship between the high-resolution images and the medium resolution ones is how we assumed. Of course, training with HD videos would obtain better results in these videos. As an important detail, one of the tested cameras does not have HD mode.
- Looking only at tests two and three, the best method is the second one. Note that it is the one that takes into account the joint information of all the images.
- All the methods improve when the number of processed frames increases, although the behavior presents ups and downs in all of them (see figure 2). Since the videos are obtained encoded in MPEG-4, a possible improvement is to extract only the images with encoding independent from the rest (intra-frame or I's images). Note that according to the curves in Figure 2 the best method is voting.

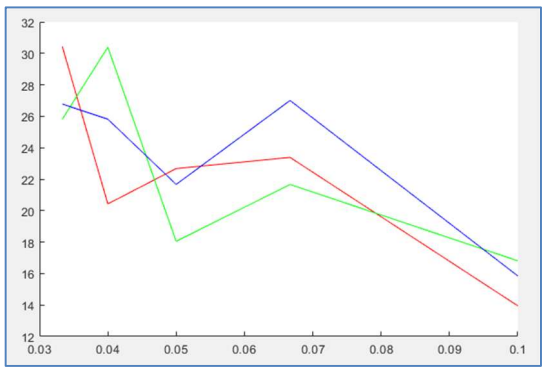

Fig. 2. Mean error (between the three tests) for the different methods versus sampling rate (1/N). Voting: red curve; PRNU pattern correlation: green curve; PCE vectors: blue curve.

More specific conclusions can be drawn by studying the individual confusion matrices (see figure 3). In general, most of the failures are concentrated in the two Logitech cameras that are the ones with the highest price and image quality. This conclusion, better sensor ➔ lower noise ➔ more difficult to identify, had already been detected in the case of camera identification with a still image [10].

```
CM1T1a =

    4    0    0    0    0
    1    5    0    0    0
    0    0    6    0    0
    1    1    1    3    0
    0    1    2    2    1

CM5T2b =

    5    0    0    0    0
    0    7    0    0    0
    0    0    7    0    0
    0    1    0    6    0
    0    0    2    0    5
```

Fig. 3. Confusion matrices for the worst case (left) and for the best (right).

## VI. CONCLUSIONS AND FUTURE LINES

Starting with a source camera identification system designed for individual images, we have developed several methods to extend this type of identification to video. We have tested them with videos obtained from Webcams, although they could be used with any video stream.

No attention was paid to the video format, although the videos used were all in MPEG-4 format, the methods could be used with any other.

The influence of the video format has not been studied, that is an interesting future line. For example, it would be interesting to study the effect of including or not in the processing the different types of images in compressed format (I's, P's and B's).

From the tested methods, the second one is the most efficient: calculation of PRNU patterns from the input videos and comparison with those stored using a correlation coefficient. All methods improve with the applied sampling rate, obtaining remarkable results with 1/10 (one image is processed out of every 10 inputs).

For effective recognition it is better, if possible, to obtain a specific collection of videos with a constant background varying different colors and, if possible, intensities.

If we have videos of different resolutions, it is better to do a training for each resolution.